\documentclass[11pt,twoside]{article}

\usepackage{asp2006}
\usepackage{epsf}
\usepackage{epsfig}
\usepackage{lscape}

\begin{document}
\title{The circumnuclear environment in M31}   
\author{Zhiyuan Li}   
\affil{Smithsonian Astrophysical Observatory, 60 Garden Street, Cambridge, MA 02138, U.S.A.}    

\begin{abstract} 
Studies of galactic circumnuclear environments
is important to our understanding of the feeding and
feedback of the central super-massive black hole (SMBH) and in turn the global evolution of the host galaxy.
We present an observational overview of the circumnuclear environment
in M31 and a tentative understanding of its regulation. Notes on
selected open issues, as well as
on a comparison with the Galactic Center and other extragalactic circumnuclear environments, are also presented.
\end{abstract}


\section{Introduction}  {\label{sec:intro}}
Galactic circumnuclear environments, in which the ISM and stars are
densely co-spatial under extreme physical conditions,
 are of vast astrophysical interest.
The Andromeda galaxy (M31) provides an ideal testbed for studying
a galactic circumnuclear environment. 
Its proximity ($D\approx780$ kpc; 1 arcmin = 230 pc) ensures
observations with an unprecedented spatial resolution for a massive
external galaxy.
Moreover, thanks to the low foreground and internal extinction, the
center of M31 is transparent in the optical, ultraviolet and X-ray
bands. 
This opens up the possibility of
a comprehensive view for nearly all phases of the ISM and various
types of stars in the region, which is inevitably hampered by the 
  edge-on perspective to our Galactic Center (GC).

Here we present an (necessarily selected) observational overview for the circumnuclear
environment in M31, with an emphasis on the multi-phase ISM and
the inherent relations among various components in this environment. 
For clarity, we loosely define the circumnuclear
region as within a galactocentric radius of 300 pc. For comparison, the
counterpart of such a region in our Galaxy is the
so-called central molecular zone (CMZ), whereas in Virgo galaxies, for instance,
the corresponding region spans an angular size of merely 3 arcsec.

\section{An observational overview}
\subsection{The nucleus} \label{subsec:nuc}

\begin{figure}[!htb]
 \centerline{
       \epsfig{figure=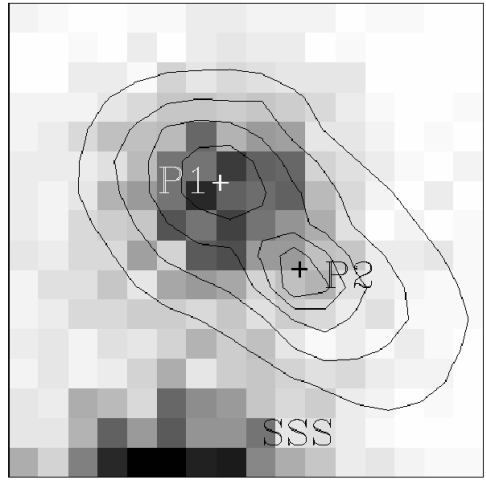,width=0.5\textwidth,angle=0}
 }
 \caption{A 0.5-8 keV {\sl Chandra}/ACIS counts image, in 0\farcs125
   pixels, overlaid by the {\sl HST}/ACS F330W intensity contours
showing the double nuclei P1 and P2. The `+' signs mark the fitted centroids of P1 and P2.
The displacement between P1 and P2 in X-ray is assumed to be same as
   in the optical. Part of a super-soft X-ray source (SSS) appears at the bottom of the image.}
\label{fig:nuc}
\end{figure}

M31 hosts the well known double optical nuclei (so-called P1
and P2; Lauer et al.~1993; see Fig.~\ref{fig:nuc}) peaking at an angular separation of about half-arcsec from each other,
which are successfully interpreted to be an eccentric disk of typically K-type stars
with a total mass of $\sim$$10^7 {\rm~M_{\odot}}$ (Tremaine 1995). 
The origin of this stellar disk is unclear. Based on their numerical
simulations, Hopkins \& Quataert (2010) suggested that a lopsided, 
parsec-scale, nuclear stellar disk can be formed in gas inflows driven by
galactic-scale instabilities and subsequently acts as a torque to fuel
the remaining
gas to the SMBH.
 
P2 is fainter than P1 in V- and B-bands
but brighter in U-band and in the near-UV. {\sl HST}/STIS spectroscopy
shows that the excess light at the shorter wavelengths can be
characterized by an A-type stellar spectrum, and is 
 consistent with a 200 Myr-old starburst embedded in P2
(dubbed a third nucleus, P3; Bender et al.~2005). 
A super-massive
 black hole (SMBH) is embedded in P2/P3, with an inferred dynamical mass of $1.4^{+0.9}_{-0.3}\times10^8 {\rm~M_{\odot}}$ (Bender et al.~2005).
Understanding star formation processes in the central parsec around
the SMBH is of astrophysical importance and currently a hotly pursued
topic in the study of our GC (this proceeding). 
Recognizing the challenge of forming stars near the SMBH, Demarque \& Virani (2007) proposed that P3  
is composed of hot horizontal branch (HB) and post-HB stars, an old
stellar origin similar to that of the
{\sl UV-upturn} phenomenon observed in
elliptical galaxies and the M31 bulge (cf.~O'Connell 1999 for a review; see \S~\ref{subsec:star}).
{\sl HST far-UV (i.e., shortward of $\sim$2000${\rm~\AA}$) spectroscopy is useful
to distinguish the young- and old-star alternatives for P3}.

Crane, Dickel \& Cowan (1992) reported the possible detection of M31$^{\ast}$ in VLA 3.6 cm observations, giving a
flux density of 28 $\mu$Jy.
Based on a 50 ks {\sl Chandra}/HRC observation, Garcia et al.~(2005)
claimed a 2.5 $\sigma$ detection of M31$^{\ast}$, the count rate of
which can be translated to a
0.3-7 keV intrinsic luminosity of $\sim9\times10^{35} {\rm~ergs~s^{-1}}$.
Basing on {\sl Chandra}/ACIS observations, Li, Wang \& Wakker (2009)
determined an intrinsic luminosity of $\sim1.2\times10^{36}
{\rm~ergs~s^{-1}}$ for P2, which sets
a firm upper limit to the (quiescent) X-ray emission from the
SMBH. The stars in P2/P3, no matter old or young, almost certainly
contribute a fraction of the observed X-ray emission, but the
extremely high stellar density there makes it difficult to reliably
quantify the stellar X-ray emission. Indeed, X-ray emission from around
the position of P1 is detected and in fact appears several times brighter than
the emission detected from around P2 (see Fig.~\ref{fig:nuc}), which Li et
al.~(2009) argued comes from one or few low-mass X-ray binaries (LMXB)
embedded in P1.   
Very recently, Garcia et al.~(2010) reported X-ray flux variation at
the levels of a factor of 3 on a timescale of days and a factor
of over 10 in a year. This variability is consistent with X-ray emission arising from
around the SMBH, but does not rule out a LMXB origin. {\sl Long-term
X-ray-radio monitoring will hopefully settle this issue}.

Compared to the SMBH in our Galaxy, Sgr A$^{\ast}$, M31$^{\ast}$ is a few times
less luminous in radio but up to three orders of magnitude more
luminous in X-ray. Both no doubt fall in the class of radiatively
inefficient accreting SMBHs (cf. Narayan \& McClintock 2008 for a review). {\sl Owing to its relatively large mass, the conventional
{\sl Bondi radius} of M31$^{\ast}$ spans an angle of
$\sim$5$^{\prime\prime}$ in the sky, opening up an intriguing
possibility of studying the spatially-resolved accretion flow}.

\subsection{Stars} \label{subsec:star}
The photometry in NIR (Beaton et al.~2007) and optical (Walterbos \& Kennicutt 1998) bands shows little color gradient
in the inner bulge. The colors are typical of an old, metal-rich stellar population, equivalent to type G5 III or K0 V.
The Mg$_2$ index of 0.324 measured from the central
$\sim$$30^{\prime\prime}$ (Burstein et al.~1988) indicates an iron metallicity
of [Fe/H]$\sim$0.3. For reference,
the K-band luminosity within the central arcmin is
4.7$\times10^{9} {\rm~L_{\odot,K}}$, which, according to the color-dependent (here a $B-V$ color of 0.95 is adopted) mass-to-light ratio of Bell \& de Jong (2001), corresponds
to a stellar mass of 4.0$\times10^{9} {\rm~M_{\odot}}$.

There is little evidence for recent massive star formation in the
circumnuclear region. Bright massive (i.e., O and B-types) stars would have
been detected at the distance of M31, but they are not observed (King et al.~1992; Brown et al.~1998).
The far-UV to near-UV
($FUV-NUV$) color in the inner bulge suggests a stellar age older than 300 Myr
(Thilker et al.~2005). While extended ionized gas is indeed present
(see \S~\ref{subsec:HII}), it shows optical line intensity ratios atypical of
conventional HII regions (Rubin \& Ford 1971; del Burgo, Mediavilla \& Arribas 2000).

The circumnuclear region of M31 has long been known to show a
UV-upturn in its spectrum (Burstein et al.~1988). {\sl HST}/FOC resolves 
the brightest UV sources, the color-magnitude diagram of which favors the
interpretation that most of them are post-HB stars (Brown et al.~2000).
The observation, however, was not deep enough to detect stars on the HB, as is
the case recently achieved for the HB stars in M32 (Brown et al.~2008). 

The stellar density in the circumnuclear region is comparable
to that in globular clusters. Hence dynamical effects are
expected to play an important role to the evolution of stars, in
particular, binaries. One such example is the LMXBs. Voss \& Gilfanov
found a significant increase 
in the specific frequency of LMXBs (i.e., per unit stellar mass)
detected within the central arcmin of M31, the radial distribution of
which is proportional to the square of stellar density. 

\subsection{Atomic gas} \label{subsec:HI}

So far there is no reported detection of atomic hydrogen in the
circumnuclear region. An upper limit of $10^6 {\rm~M_{\odot}}$ is set
on the HI mass within the central 500 pc (Brinks 1984). Results of a recent
wide-field HI imaging survey of M31 (Braun et al.~2009) seem to be
compatible with this value (Braun 2009, private communication).

\subsection{Warm ionized gas} \label{subsec:HII}
The existence of ionized gas has long been known through the detection
of [O II], [O III], H${\alpha}$, [N II] and [S II] emission lines in
the spectra of the inner bulge (M${\rm \ddot{u}}$nch 1960; Rubin \&
Ford 1971). Later narrow-band imaging observations (Jacoby, Ford \&
Ciardullo 1985; Ciardullo et al.~1988; Devereux et al.~1994) further
revealed that the gas is apparently located in a thin plane, showing
filamentary and spiral-like patterns, across the central few arcmins
(so-called a {\sl nuclear spiral}; see Fig.~\ref{fig:ha_ir}).
The electron density of the ionized gas, inferred from the intensity ratio of [S II] lines, is $\sim$$10^2$-$10^4 {\rm~cm^{-3}}$ within the central arcmin, generally decreasing
outward from the center (Ciardullo et al.~1988). The gas is estimated to have a mass of $\sim$$10^3{\rm~M_{\odot}}$, an H${\alpha}$+[N II] luminosity of a few $10^{39} {\rm~ergs~s^{-1}}$, and a very low volume filling factor consistent with its filamentary morphology (Jacoby et al.~1985).
The relatively high intensity ratio of [N II]/H${\alpha}$, ranging from $\sim$1.3-3 in different regions (Rubin \& Ford 1971; Ciardullo et al.~1988), is similar to the typical values
found in early-type galaxies (e.g., Macchetto et al.~1996) rather than in conventional HII regions (where typically [N II]/H${\alpha}$ $\sim$0.5).
The kinematics of the gas is rather complex. A major component of the velocity field apparently comes from
circular rotation, whereas the residuals indicate both radial and vertical motions (Rubin \& Ford 1971).

That the stellar disk of M31 is probably barred (Athanassoula \&
Beaton 2006) offers a natural formation mechanism for the nuclear spiral: an inflow of gas from the outer disk
driven by bar-induced gravitational perturbations to form organized patterns (e.g., Maciejewski 2004).
Indeed, by modelling the gas dynamics in a bar-induced potential Stark \& Binney (1994) obtained a satisfactory fit to the observed position-velocity diagram of the ionized and neutral gas in the central $\sim$2$^\prime$.
Another possible driver of gas is a recent head-on collision between M31 and its companion galaxy, most likely M32 (Block et al.~2006).
Although details remain to be studied, it seems certain that an asymmetric gravitational potential is responsible for
the formation and maintenance of the nuclear spiral in M31, and perhaps so for similar gaseous structures found in the inner regions of disk galaxies.
The CMZ in our GC, for example, is thought to be formed in
such processes (cf. Morris \& Serabyn 1996 for a review).

The ionizing source of the nuclear spiral remains uncertain, however,
especially in view of the lack of massive stars {\sl in situ}. 
Li et al.~(2009) considered possible alternative sources for the
ionizing photons (i.e., shortward of 912${\rm~\AA}$), which include: i) UV radiation
of hot evolved stars, such as post-AGB stars and HB stars; ii) X-ray
photons from the hot gas (see \S~\ref{subsec:hot}) as well as stellar objects; iii)
UV photons induced by thermal conduction ({\S~\ref{sec:dis}); and
iv) UV photons induced by interstellar shocks;
Li et al.~(2009) drew a tentative conclusion that the only likely ionizing source is the stellar UV radiation, predominantly
contributed by post-AGB stars, with an additional contribution
from HB stars. The other sources considered all fall short of accounting for the required ionizing photons.
It is worth noting that their estimates of the ionizing flux from post-AGB
and HB stars are based on synthesis stellar evolution models
considered by Binette et al.~(1994) and Han et al.~(2007). The
adequacy of such models are challenged by recent {\sl HST} far-UV 
observation of the post-AGB and HB populations in M32 (Brown et al.~2008).  
{\sl Therefore more accurate estimates should be based on direct counting
of such objects in the vicinity of the nuclear spiral, which requires
{\sl HST} far-UV imaging of deeper exposure and wider spatial coverage}.

Moreover, it was realized that photoionization models for post-AGB stars predict an [N II]$\lambda$6584/H$\alpha$ intensity ratio of $\sim$1.2
for gas with an abundance up to 3 solar (Binette et al.~1994), which is
inconsistent with the high ratios (generally $>$1.3, as large as
$\sim$2.7) observed in M31 (Ciardullo et al.~1988).
Not to root on anomalously high nitrogen abundance,  
the relatively high [N II]/H$\alpha$ ratio implies that heating in addition to photoionization contributes substantially to the production of the nitrogen ions.
The heating source remains to be identified. A likely candidate is high-energy particles
produced by the SMBH or supernova (SN) events (see \S~\ref{subsec:mag}).

An additional interesting remark arises for narrow optical line emission
 detected in the core of more distant galaxies, especially those
 dubbed low-ionization nuclear emission-line regions (LINERs; cf. Ho
 2008 for a review).  LINERs observed in the Palomar Survey (Ho, Filippenko \&
 Sargent 1997) have a median H$\alpha$ luminosity of $\sim$$2\times
 10^{39}{\rm~ergs~s^{-1}}$, a value comparable to that of 
 the nuclear spiral in M31. The observed [N II]/H$\alpha$ ratio in M31
 is also similar to those of LINERs. 
 That neither M31$^\ast$ nor on-going star
 formation is likely responsible for photon-ionizing the nuclear
 spiral presents interesting implications to the
 ionization/excitation mechanisms of LINERs. 

\begin{figure*}[!htb]
\vskip -1cm
 \centerline{
  \epsfig{figure=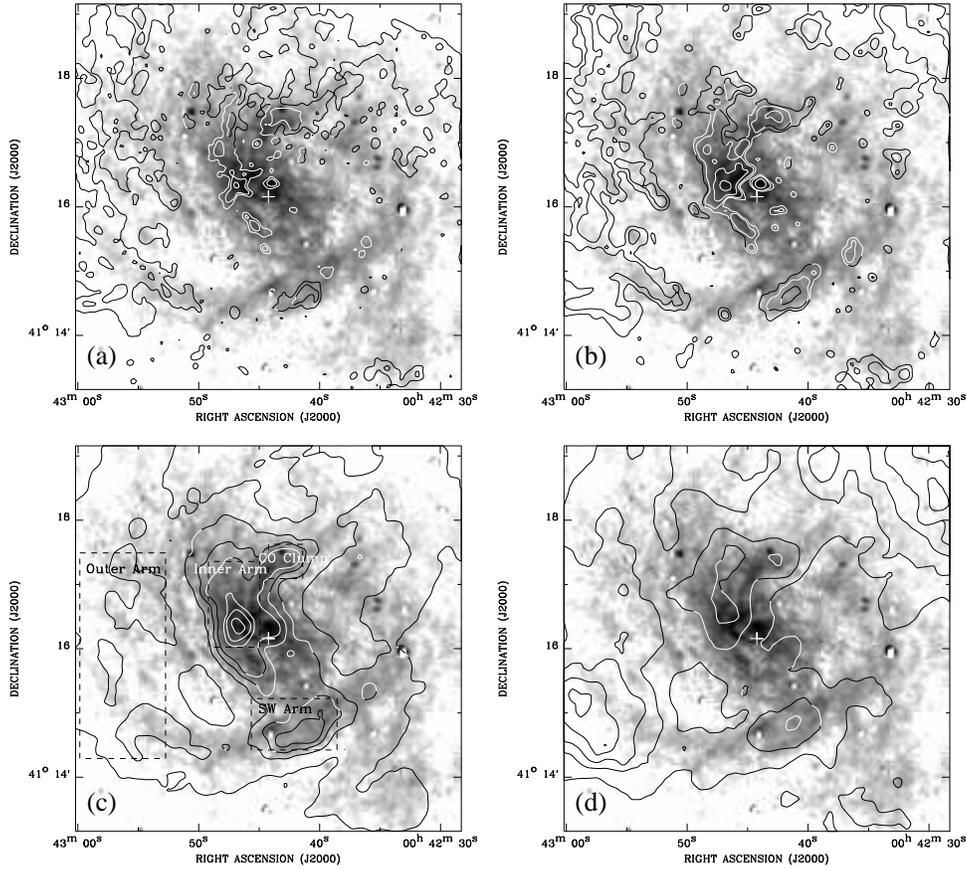,width=\textwidth,angle=0}
 }

   \caption{Contours of (a) 8 $\mu$m, (b) 24 $\mu$m, (c) 70 $\mu$m and
     (d) 160 $\mu$m emission overlaid on the H${\alpha}$ image
     (Devereux et al.~1994)
of the central $6^\prime$ by $6^\prime$ region, in arbitrary units,
     showing the nuclear spiral.
The dashed rectangles
marked in (c) outline selected regions of interest.
   }
 \label{fig:ha_ir}
\end{figure*}

\subsection{Dust and molecular gas} \label{subsec:dust}
Probing circumnuclear dust in M31 via optical extinction has a long history (e.g., Johnson \& Hanna 1972; Sofue et al.~1994).
{\sl Spitzer} observations now provide the highest-resolution
mid-IR/far-IR (MIR/FIR)
view toward M31 (Barmby et al.~2006; Gordon et al.~2006), in particular to its circumnuclear regions. By
properly subtracting stellar IR emission, Li et al.~(2009) obtained
the IR emission of circumnuclear dust, the morphology of which
markedly resembles that of the optical emission lines, i.e., the
nuclear spiral (see Fig.~\ref{fig:ha_ir}). 

Molecular gas in the circumnuclear region remains poorly studied
to date. Detection of CO closest to the M31 center ($\sim$1\farcm3 away) points to a prominent dust complex, D395A/393/384, with an estimated molecular gas
mass of 1.5$\times10^{4} {\rm~M_{\odot}}$ (Melchior et al.~2000).
This 100 pc-wide feature is also seen in the MIR/FIR emission (Fig.~\ref{fig:ha_ir}c).
The total mass of molecular gas in the circumnuclear region of M31 is
estimated to be $\sim$$10^{6} {\rm~M_{\odot}}$ (Li et al.~2009),
roughly an order of magnitude lower than that of the CMZ.


\subsection{X-ray-emitting hot gas} \label{subsec:hot}
While X-ray emission from M31 has been detected for more than three
decades (Bowyer et al.~1974), the bulk of this emission is thought to
arise from X-ray binaries. Only recently, and thanks to the superb angular
resolution and sensitivity of {\sl Chandra} observations, has the presence of
X-ray-emitting diffuse hot gas in the M31 bulge been unambiguously 
confirmed (Li \& Wang 2007; Bogd\'{a}n \& Gilfanov 2008).
Morphologically, the diffuse X-ray emission appears
elongated approximately along the minor-axis of the disk (see
Fig.~\ref{fig:dif}a), reminiscent of a bi-polar outflow from the bulge.

\begin{figure*}[!htb]
  \centerline{
    \epsfig{figure=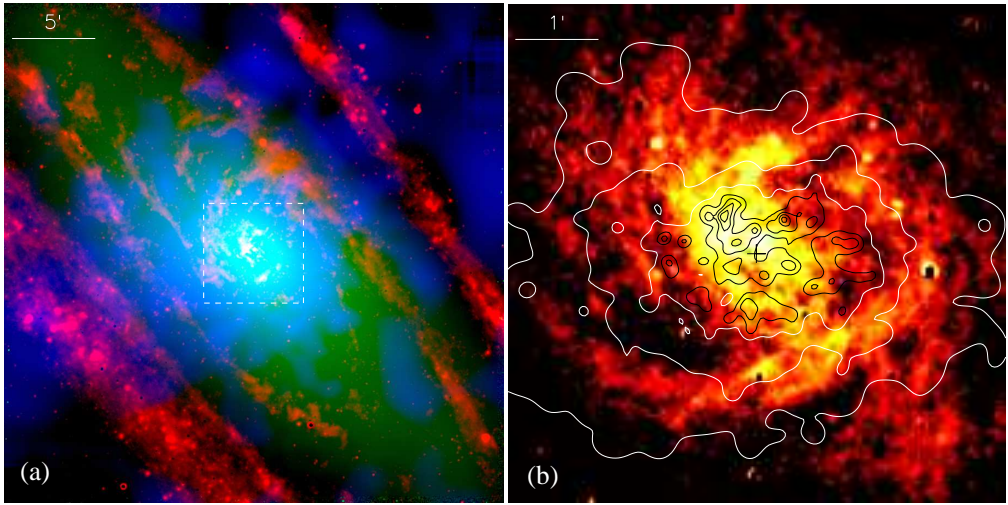,width=\textwidth,angle=0}
  }
  \caption{(a) Tri-color image of the central 30$^\prime$ by 30$^\prime$ (6.8 kpc by 6.8 kpc) of M31. {\sl Red}: {\sl Spitzer}/MIPS 24 $
\mu$m emission; {\sl Green}: 2MASS K-band emission; {\sl Blue}: {\sl Chandra}/ACIS 0.5-2 keV emission
of diffuse hot gas. The dashed box outlines the central
6$^{\prime}$ by 6$^{\prime}$,  a region further shown in (b) and Fig~\ref{fig:ha_ir}. (b) Smoothed intensity contours of the 0.5-2 keV diffuse emission overlaid on the H${\alpha}$ emission. 
The plus sign marks the M31 center. North is up.}
 \label{fig:dif}
\end{figure*}

With the advance of the high-resolution {\sl Chandra} view
to the circumnuclear region (Li et al.~2009; see Fig.~\ref{fig:dif}b), interesting
relations between the hot gas and the nuclear spiral are
revealed. The diffuse X-ray emission appears stronger on the southeastern
side (i.e., along the minor-axis), consistent with it being absorbed by the tilted nuclear spiral, if
the emission is intrinsically quasi-symmetric. In the very central region,
the X-ray emission peaks near where the H$\alpha$ emission peaks. 
The {\sl Chandra} spectrum of the hot gas can be characterized by a
single temperature ($\sim$0.3 keV) plasma model (Li et al.~2009). Assuming 
that the hot gas fills the bulk circumnuclear volume, the gas density
is estimated to be $\sim$$0.1{\rm~cm^{-3}}$. {\sl XMM-Newton}/RGS
spectrum of higher spectral
resolution further reveals 
an oxygen-to-iron abundance ratio of $\sim$0.3 solar (Liu et
al.~2010), indicating metal-enrichment by SNe Ia.

It should be noted that such a 0.3 keV gas, even if it exists, would not be
detected for the GC due to the heavy foreground absorption.  
While the presence of hot gas in the GC is probably beyond doubt,
especially in the inference of M31, its physical properties 
are currently quite uncertain (this proceeding).

\subsection{Magnetic field and high energy particles} \label{subsec:mag}
At $10^{\prime\prime}$-$30^{\prime\prime}$ resolution the radio continuum emission shows filamentary patterns apparently associated with the H${\alpha}$ emission, i.e., the nuclear spiral (Walterbos \& Grave 1985; Hoernes, Beck \& Berkhuijsen 1998).
The average power-law spectral index of $\alpha$$\sim$-0.75 ($S_\nu \propto \nu^\alpha$) throughout the 2.8-73.5 cm wavelength range indicates that the bulk emission is non-thermal (Walterbos \& Grave 1985).
Assuming energy equipartition and a volume filling factor of unity
(which is probably substantially overestimated), Hjellming \& Smarr (1982)
estimated an energy density of $\sim$$0.5{\rm~eV~cm^{-3}}$ for the energetic particles within the central 30$^{\prime\prime}$.
In principle, both the SMBH and the SN Ia events in the bulge can
produce copious high energy particles. However, radio observations of
the circumnuclear region of M31 remain poorly investigated, although they are
no doubt crucial to shed light on the nature of high energy
particles and magnetic field.

   
\section{Regulation of the circumnuclear environment} \label{sec:dis}
A general picture of the circumnuclear environment in M31 is presented in the above. In the central few hundred parsecs, the ISM consists of two dynamically distinct components. One is the nuclear spiral with a low volume filling factor,
consisting of cold dusty gas, traced by the MIR and FIR emission, and warm ionized gas, traced by optical recombination lines.
The nuclear spiral is thought to be formed by bar-induced gravitational perturbations with a possibly continuous supply of gas from the outer disk regions.
Connections between the nuclear spiral and the major spiral arms in the outer disk are evident in Fig.~\ref{fig:dif}a.
The other component is a corona of volume-filling hot gas, traced by the diffuse X-ray emission. This hot corona
has a bi-polar extent of at least several kpc
away from the midplane.
While young massive stars are essentially absent, embedded in the hot corona is an old stellar population with a total mass of $\sim$$10^{10}{\rm~M_\odot}$, which is primarily responsible for the gravitational potential and probably for the energetics of the ISM.
Finally, there is the radiatively inactive SMBH manifesting itself
only in radio and X-ray to date.
Both the circular speed of the disk ($v_c\sim270 {\rm~km~s^{-1}}$ at $r\approx230 {\rm~pc}$) and the
sound speed of the hot gas ($c_s\sim280 {\rm~km~s^{-1}}$ at a temperature of 0.3 keV)
imply a relatively short dynamical timescale of $\sim$$10^6$ yr.
Unless our multiwavelength view is a highly transient one, which is unlikely, there ought to be 
certain physical processes regulating the behavior of the multi-phase ISM as well as that of the SMBH.

Li et al.~(2009) proposed a scenario for such a 
regulation, and we simply outline the basic idea here.
Commonly suggested for intermediate-mass, relatively
isolated, early-type galaxies, diffuse hot gas in such systems arises from 
the collective mass loss of evolved stars (e.g., stellar winds, planetary
nebulae) heated by
SNe Ia (e.g., Ciotti et al.~1991; David et al.~2006; Li, Wang
\& Hameed 2007; Tang et al.~2009).
This should be the case in the M31 bulge, where the SMBH is quiescent
and there is no recent massive star formation.
In the present-day universe, a stellar spheroid empirically deposits energy and mass
at rates of $\sim$$1.1\times10^{40}$$[L_K/(10^{10} L_{\odot,K})]$${\rm~ergs~s^{-1}}$ (Mannucci et al.~2005)
and $\sim$$2{\times}10^{-2}[L_K/(10^{10} L_{\odot,K})]$${\rm~M_\odot~yr^{-1}}$ (Knapp, Gunn \& Wynn-Williams 1992), respectively,
where $L_K$ is the K-band luminosity and a proxy of stellar mass. Assuming a unity SN heating efficiency and that the stellar mass loss is
wholly thermalized, an
average energy input of $\sim$3.6 keV per gas particle is inferred. The gravitational potential of a normal galaxy
is unlikely to confine the gas with such a high temperature. Therefore the gas is expected to escape, at least from inner regions
of the host galaxy. Indeed, for many early-type galaxies, including M31, the observed X-ray luminosity of hot gas is typically no more than a few
percent of the expected energy
input rate from SNe Ia (e.g., David et al.~2006; Li, Wang \& Hameed 2007); the inferred gas mass is also
much less than expected if the stellar ejecta has been accumulating for a substantial fraction of the host galaxy's lifetime.
Such discrepancies can be explained naturally with the presence of an outflow of hot gas, in which the ``missing'' energy and mass
are transported outside the regions covered by the observations.
On the other hand, the gas temperature in the M31 bulge, $\sim$0.3 keV, is much lower than
the maximum allowed temperature of $\sim$3.6 keV,
a pressing discrepancy again often noted for early-type galaxies.
Taking into account interaction between the nuclear spiral and the hot
gas helps solve this
discrepancy. Specifically, {\sl thermal conduction} serves to
evaporate the cold gas, turning it into, and effectively lowering the
temperature of, the hot phase.
Such a process naturally leads to (1) the starving of M31$^\ast$ and
the absence of active star formation, in
spite of a probably continuous supply of gas from outer disc regions and (2) the launch of a
bulge outflow of hot gas, primarily mass-loaded from the circumnuclear
region. 

The above scenario, albeit crude in details, is self-consistent in
view of the current multiwavelength information (cf. Li et al.~2009
for detailed discussions).
One particular
prediction of such a scenario is the presence of gas with intermediate
temperatures arising from the conductive interfaces.
Indeed, Li et al.~(2009) found tentative evidence of such, 
including enhancement of both soft X-ray emission and FUV
emission near the nuclear spiral. 
A significant charge exchange contribution to the OVII triplet, which
can be attributed to interactions between highly ionized gas and
neutral gas, is also hinted in the {\sl XMM-Newton}/RGS spectrum (Liu
et al.~2010).
Moreover, an effective evaporation of the nuclear spiral requires that
it consist primarily of individual small gas clouds of sub-parsec sizes,
such that the mean free path
of conducting electrons is shorter than the scale-length of temperature variation.
This is supported by the
{\sl HST} continnuum images, in which the nuclear spiral
clearly manifests itself as extinction features against the bulge starlight,
exhibiting a variety of fine structures down to the image
resolution of $\sim$0\farcs1 (Z.~Li et al.~in preparation).
{\sl More conclusive tests await
{\sl HST} imaging spectroscopy along the nuclear spiral (e.g., those selected
regions outline in Fig.~\ref{fig:ha_ir}c) to probe the spatial and
kinematic information of  
optical and FUV line emission  on sub-parsec
scales}.

The scenario is potentially applicable to other galactic circumnuclear environments,
particularly those in early-type spirals with a substantial
bulge. It is also reasonable to invite its further application to elliptical/lenticular
galaxies, in which hot and cold gas are often observed to co-exist
and sometimes show morphological correlations (e.g. Trinchieri,
Noris \& di Serego Alighieri 1997; Sarzi et al.~2010). In any
case, thermal conduction is expected to be prevalent in the core of
early-type galaxies containing typically a dense, multi-phase ISM, a
process often overlooked.

\section{Summary}
M31 provides the nearest testbed for a
multiwavelength study of a circumnuclear environment, which, in the
author's point of view, serves as a benchmark for studying
galactic circumnuclear environments in general.

\acknowledgements 
I am grateful to my collaborators Michael Garcia, Jiren Liu, Bart Wakker and Daniel Wang.


\end{document}